\documentclass[letterpaper,12pt]{article}
\usepackage{amsmath}%
\usepackage{amsfonts}%
\usepackage{amssymb}%
\usepackage{hyperref}%
\usepackage{epsf}%

\begin{document}
\begin{center}
{ \LARGE {\bf The quantum gravitational black hole}}\\
{ {\LARGE {\bf is neither black nor white}}}
\vskip 2cm
{\bf {\large
{\bf T. P. Singh$^{a}$\footnote{e-mail address: tpsingh@tifr.res.in} and
Cenalo Vaz$^{b}$\footnote{e-mail address: vaz@physics.uc.edu}
}}}
\vskip 1cm
{\it $^{a}$Tata Institute of Fundamental Research,}\\
{\it Homi Bhabha Road, Mumbai 400 005, India.}
\vskip 0.5cm
{\it $^{b}$Department of Physics, University of Cincinnati}\\
{\it Cincinnati, Ohio, USA}\\
\end{center}
\vskip 1cm
\begin{abstract}
\bigskip

\noindent Understanding the end state of black hole evaporation,
the microscopic origin of black hole entropy, the information loss paradox, and
the nature of the singularity arising in gravitational collapse - these are
outstanding challenges for any candidate quantum theory of gravity. Recently,
a midisuperspace model of quantum gravitational collapse has been solved using
a lattice regularization scheme. It is shown that the mass of an eternal
black hole follows the Bekenstein spectrum, and a related argument
provides a fairly accurate estimate of the entropy. The solution also
describes a quantized mass-energy distribution around a central black hole,
which in the WKB approximation, is precisely Hawking radiation. The leading
quantum gravitational correction makes the spectrum non-thermal,
thus providing a plausible resolution of the information loss problem.
\end{abstract}
\newpage

\baselineskip 19pt

Recent developments in quantum general relativity have provided important
insights into long-standing questions relating to the end-state of  black hole
evaporation,  the origin of black-hole entropy, the information
loss paradox and the fate of the singularities arising in classical gravitational
collapse. These advances in our understanding of collapse have become possible
because of a successful canonical quantization and lattice regularization of a
midisuperspace model of quantum gravitational collapse \cite{vazrev}. They are in
keeping with the new understanding that has recently been achieved with regard to the
cosmological singularity in loop quantum cosmology \cite{bojowald}.

Following earlier pioneering work by Kucha\v r \cite{kuc1} and by Kastrup and Thiemann
\cite{kt1}, we developed a canonical description of spherical dust collapse as
described by the Lemaitre-Tolman-Bondi models.  For the marginally bound models, the
canonical variables are the area radius $R$, the dust proper time $\tau $, the mass function,
$F=2M$, of the collapsing dust cloud and their conjugate momenta. The momentum
conjugate to $F$ can be eliminated using the momentum-constraint, to obtain the following
simple form for the Hamiltonian constraint \cite{vws}
\begin{equation}
\label{ham}(P_\tau +F^{\prime }/2)^2+{\cal F}P_R^2-\frac{F^{\prime 2}}{4%
{\cal F}}=0,
\end{equation}
where $F^{\prime }$ is the derivative of $F(r)$ with respect to the
comoving coordinate $r$ (it represents the shell mass density), and ${\cal F}=1-F/R$.

 One can introduce a flat DeWitt metric on the configuration space by defining
$R_{*}=\int dR/\sqrt{|{\cal F}|}$. Then the quantum gravitational collapse corresponding
to (\ref{ham}) is described by the Wheeler-DeWitt equation
\begin{equation}
\label{wd}\left[ \frac 1{c^2}\frac{\delta ^2}{\delta \tau ^2(x)}\pm \frac{%
\delta ^2}{\delta R_{*}^2(x)}\pm \frac{f^{\prime 2}}{4l_p^2|{\cal F}|}%
\right] \Psi [\tau ,R]=0,
\end{equation}
where we have used the dimensionless variables $x=r/l_p$, and $f(x)=F/l_p$. The upper sign
refers to the region $R>F$, and the lower sign to $R<F$.  The apparent horizon is the
curve $R=F$. The momentum constraint is
satisfied if the wave functional is a spatial scalar, which we assume to be of the
``stationary state'' form
\begin{equation}
\label{wf}\Psi [\tau ,R]=\exp \left[ -\frac{ic}{2l_p}\int_0^\infty
dxf^{\prime }(x)(\tau +{\cal U}(R))\right] .
\end{equation}

To regularize the functional derivatives, we choose a lattice by dividing space into
cells, the size of the $j^{\rm th}$ cell being $\sigma _j$, and we finally take the limit
$\sigma_j\rightarrow 0,\forall j$. Remarkably enough, the lattice size drops out of the
Wheeler-DeWitt equation, which factors into an infinite set of ordinary differential equations
for the time independent wave functions, one for each cell \cite{vws2}:
\begin{equation}
\label{ode}z(z-1)^2\frac{d^2y}{dz^2}+\frac{z-1}2\frac{dy}{dz}+\gamma
^2z^2y=0.
\end{equation}
The quantities $y,z$ and $\gamma $ are defined independently for each cell, and for the
$j^{th}$ cell are given by $z_j=R_j/F_j$, $\gamma _j=F_j\omega_j/c$ and $y_j\equiv \Psi _j(z)$,
which is the time-independent wave function for the $j^{th}$ cell. The frequency, $\omega _j$,
defined by the relation
\begin{equation}
\label{fre}f_{j+1}-f_j\equiv \frac{2l_p}c\omega _j,
\end{equation}
is related to the energy density, $\varepsilon_j$, of the $j^{\rm th}$ shell by
$\varepsilon_j =\hbar \omega_j$.

The simple form to which the canonical quantum dynamics of the collapsing cloud has been
reduced pays off rich dividends. One is able to draw significant conclusions about the
Bekenstein mass spectrum, black hole entropy, and quantum gravitational corrections to
Hawking radiation.

An eternal black hole is described by a mass function $F(r)$ which is non-vanishing only
at the origin (we recall that the Schwarzschild black hole is a special case of the
Tolman-Bondi solution). In this case, the above midisuperspace problem reduces to quantum
mechanics. The stationary states of the black hole are a superposition of ingoing and outgoing
waves in the interior and in the exterior they are exponentially decaying  because the
Wheeler-Dewitt equation is elliptic when $R>F$. Matching the wave function and its derivative
at the horizon directly yields the Bekenstein mass
spectrum \cite{bek}
\begin{equation}
\label{bek}M^2=\left( n+\frac 12\right) M_p^2.
\end{equation}
Insight into the origin of black hole entropy can be obtained by taking a mass function
$F=2M$ that describes a series of successive collapsing shells, each of which obeys
a mass quantization condition analogous to (\ref{bek}),
\begin{equation}
\mu_j M_j = \left(n_j + \frac{1}{2}\right) M_p^2
\label{shellcond}
\end{equation}
where $\mu_j$ is the mass of the $j^{\rm th}$ shell and $M_j$ is the mass contained within it.
These conditions, when applied recursively, show that the mass of the $j^{\rm th}$ shell is
determined by $j$ quantum numbers. Thus the total mass of a quantum black hole formed by $N$
collapsed shells will depend on $N$ quantum numbers and a quantum black hole cannot be described
simply by its total mass because such a description would ignore the manner in which the mass
was distributed among the shells. The entropy counts the number of distributions for a given
total mass. For an {\it eternal} black hole, $M_k$ in (\ref{shellcond}) should be replaced by
$M$, the mass of the hole. The total mass (squared) of the hole
continues quantized as before and the problem of counting the number of distributions is
precisely the problem of asking for the number of ways in which $N$ integers may be added
to give another integer. This result depends on the number of shells that have collapsed
to form the black hole, which we do not know but which can be independently determined by
maximizing the entropy with respect to $N$. When both $N$ and $M/M_p$ are large, one readily
finds, to leading order \cite{vw1},
\begin{equation}
S \approx 0.962 \times \frac{\mathcal A}{4}
\end{equation}
in units of Planck area, which agrees well with the Bekenstein-Hawking value.

In order to describe Hawking radiation \cite{haw1}, we assume the mass function to be a monotonically
increasing function of $x=r/l_p$. This then encodes information about the black hole as well as
the emitted radiation, and Eqn. (\ref{ode}) is to be solved for a given shell, with some assumed
mass function. One is interested in the asymptotic solution as $z\rightarrow \infty ,$ and in
the near horizon, $z=1$, solution to this equation. The asymptotic solution which describes
outgoing waves on ${\cal I}^{+}$ is found to be
\begin{equation}
\label{asy}\Psi ^\infty =\prod_je^{-\frac{i\epsilon _j}\hbar \left( \tau _j-%
\frac{2F_j}c\sqrt{z_j}+\frac{i\hbar }{4F_j}\ln z_j\right) }
\end{equation}
($\epsilon _j=\hbar \omega _j),$ and the near horizon solution which describes outgoing shells
scattered near their horizons is
\begin{equation}
\label{hor}\Psi ^{hor}=\prod_je^{-\frac{i\epsilon_j }\hbar \left( \tau _j+%
\frac{F_j}c\ln |z-1|+\frac{i\hbar }{4\epsilon _j}\ln |z_j-1|\right) }.
\end{equation}
If in these two wave functionals the last, $\hbar -$ dependent, term is dropped from the exponent,
they coincide exactly with the WKB approximation. Hawking radiation is inferred by calculating
the shell by shell Bogoliubov coefficient
\begin{equation}
\label{bet}\beta (\omega ,\omega ^{\prime })=<\Psi _{\omega ^{\prime
}}^{\infty \dagger }|\Psi _\omega ^{hor}>
\end{equation}
for the WKB wave functionals, giving the expected result
\begin{equation}
\label{bsq}|\beta (\omega ,\omega ^{\prime })|^2=2\pi ^2F^2\frac{kT_H}%
\varepsilon \frac 1{e^{\frac \varepsilon {kT_H}}-1}
\end{equation}
which is the Hawking spectrum at the Hawking temperature $kT_H=\hbar c^3/8\pi GM$. An exact series
solution to (\ref{ode}) gives the same result \cite{vws2}. The $\hbar -$dependent term in (\ref{hor})
is responsible for modifying the Hawking radiation, and a fresh calculation of the spectrum with
this term retained gives the corrected relation
\begin{equation}
\label{cor}|\beta (\omega ,\omega ^{\prime })|^2=2\pi ^2F^2\frac{kT_H}%
\varepsilon \frac 1{e^{\frac \varepsilon {kT_H}}-1}\sqrt{\frac{2c}{F\omega
^{\prime }}}\left[ 1-\frac 12\ln \left( \frac{\pi kT_H}\varepsilon \right)
\right] .
\end{equation}
This correction cannot be obtained by modifying the Hawking temperature, and renders the radiation
non-thermal. Also, its not equivalent to a correction to black hole entropy - corrections to entropy
have been computed by various approaches in the past, but they can all be understood as relating to
a thermal spectrum, simply modifying the Hawking temperature. The non-thermal correction suggests
that unitarity may not break down in quantum gravitational evolution, and information is not lost.
Such a conclusion is also suggested by examining the history of a collapsing shell, as described in
\cite{haj1,haj2}.

The formalism for analysing quantum gravitational collapse presented here will be of significance
also for a proper understanding of the naked singularities that arise in spherical classical
gravitational collapse. A semiclassical picture for the evaporation of a naked singularity turns
out to be inadequate. This is because the outgoing quantum flux as seen by an asymptotic observer
diverges on the Cauchy horizon. A closer examination, however, reveals that this semiclassical
approximation breaks down very early during the collapse leading to a naked singularity \cite{six}
(for astrophysically relevant collapse, this time scale is the collapse time scale - typically of
the order of milli-seconds).

This breakdown occurs because the central curvature very quickly approaches Planck scales, beyond
which the semiclassical picture is no longer valid. It is also found \cite{six} that the collapsing
star emits a negligible amount of energy (of the order of a Planck unit) during the semiclassical
phase. Thus essentially the entire star enters its quantum gravitational phase, on a very short time
scale,  without any significant Hawking emission. This behaviour should be contrasted with that of
an evaporating astrophysical black hole, whose semiclassical evaporation phase is much longer than
the age of the universe, and the quantum gravity epoch is reached only after the entire star has
nearly evaporated. A naked star is thus the first system known to physicists whose complete evolution
cannot be understood without quantum gravity, and which reaches its quantum gravitational phase very
much within the lifetime of the universe. Whether quantum gravity causes the naked star to explode
catastrophically, or whether it settles down into a gentle black hole state, is a question the answer
to which is in fact contained in the Wheeler-DeWitt equation (\ref{wd}).

Classical black holes are black, semiclassical black holes are white, but the non-thermal correction
suggests that the quantum gravitational black hole is more like grey. (These grey body factors are
associated with the horizon, and are completely different from the grey body factors associated
with the radiation at infinity; the latter originate in the back-scattering against a classical
spacetime geometry.) Hawking's pioneering work on black hole evaporation left us with essentially
three options, {\it viz.,} (a) information is indeed lost during the process {\it i.e.,} the
evolution is not unitary, or (b) information is not lost, the semiclassical treatment is misleading
and the full quantum evolution is in fact unitary, or (c) a remnant that carries
with it all the information content of the black hole is left behind. Our result can be seen as making the case
for the second and/or third of the above options much stronger. Moreover, our methods can also be used
as the starting point for a  serious look at quantum gravity as the Cosmic Censor.

\end{document}